\documentclass[11pt]{article}
\usepackage[normalem]{ulem}
\pdfoutput=1 

\usepackage{jheppub} 
\usepackage{url}
\usepackage{caption}
\usepackage{subcaption}
\usepackage{extarrows}

\usepackage[latin9]{inputenc}
\usepackage{float}
\usepackage{amsmath}
\usepackage{amssymb}
\usepackage{graphicx}
\usepackage{esint}
\usepackage{hyperref}
\usepackage{comment}
\usepackage{color}
\usepackage{microtype}
\usepackage{cleveref}
\usepackage{breakurl}
\usepackage{bbm}
\newcommand{\be}{\begin{equation}}
\newcommand{\ee}{\end{equation}}
\newcommand{\ben}{\begin{displaymath}}
\newcommand{\een}{\end{displaymath}}
\newcommand{\bea}{\begin{eqnarray}}
\newcommand{\eea}{\end{eqnarray}}
\def\K{K{\"a}hler }
   \newcommand{\rf}[1]{(\ref{#1})}
\newcommand{\vp}{\varphi}

\def\be{\begin{equation}}
\def\ee{\end{equation}}
\def\bea{\begin{eqnarray}}
\def\eea{\end{eqnarray}}
\def\ba{\begin{array}}
\def\ea{\end{array}}
\def\bit{\begin{itemize}}
\def\eit{\end{itemize}}

\def\a{\alpha}

\def\vp{\varphi}

 \makeatletter

\allowdisplaybreaks

\makeatother

\makeatletter
\DeclareRobustCommand{\rcite}[1]{%
  \rcite@aux#1,\@nil{#1}%
}
\def\rcite@aux#1,#2\@nil#3{%
  \if\relax#2\relax
    Ref.~\cite{#3}%
  \else
    Refs.~\cite{#3}%
  \fi
}
\makeatother

\hypersetup{
    colorlinks = true,
    citecolor = {blue},
    linkcolor = {blue},
    urlcolor = {blue},
}

 \title{\rm { \huge \bf   Hybrid  cosmological attractors }}
 
\author{Renata Kallosh and }
\author{Andrei Linde}

\affiliation{Stanford Institute for Theoretical Physics and Department of Physics,\\ Stanford University, Stanford, CA 94305, USA}
\emailAdd{kallosh@stanford.edu}
\emailAdd{alinde@stanford.edu}

\abstract{ 
We construct $\alpha$-attractor versions of hybrid inflation models. In these models, the potential of the inflaton field $\varphi$ is  uplifted by  the potential of the second field $\chi$. This uplifting ends   due to a tachyonic instability with respect to the field $\chi$,  which appears when $\varphi$ becomes  smaller than some critical value $\varphi_{c}$. In the large $N$ limit, these models have the standard universal $\alpha$-attractor predictions. In particular,   $n_{s }= 1-  {2  \over  N}$ for the exponential attractors. However,  in some special cases the large $N$ limit is reached only beyond the horizon, for $N \gtrsim  60$. This may change  predictions for the cosmological observations. For any fixed $N$, in the limit of large uplift $V_{\rm up}$, or in the limit of large $\varphi_{c}$,  we find another attractor prediction,  $ n_s = 1$. By changing the parameters $V_{\rm up}$ and $\varphi_{c}$ one can continuously interpolate between the  two attractor predictions $n_{s }= 1-  {2  \over  N}$ and $n_{s}  = 1$. This provides significant flexibility, which can be very welcome in view of the rapidly growing amount and precision of the cosmological data. Our main result is not specific to the hybrid inflation models. Rather,  it is generic to any inflationary models where the inflaton potential, for some reasons, is uplifted, and  inflation ends prematurely.}

\begin{document}

\maketitle


\parskip 4pt

\section{Introduction}

In this paper we will study two-field cosmological attractors, using the $\alpha$-attractor generalization of the original version of hybrid inflation as an example \cite{Linde:1991km,Linde:1993cn}.

In cosmological $\a$-attractors of a single inflaton field, the predictions for the spectral index $n_s$ and for the tensor to scalar ratio $r $ are very stable with respect to significant modifications of the inflaton potential. The inflaton field  in these models can be real, but the most interesting interpretation of these models appears in supergravity describing complex fields with hyperbolic geometry  \cite{Kallosh:2013hoa,Ferrara:2013rsa,Kallosh:2013yoa,Galante:2014ifa,Kallosh:2015zsa,Kallosh:2019hzo}. In such models, kinetic terms of the scalar field   are singular at the boundary of the hyperbolic space. The singularity disappears after a transformation making the real part of the scalar field   canonically normalized. This transformation modifies the original inflaton potential $V$, which acquires an infinitely long plateau in terms of the canonically normalized inflaton field $\vp$.

 In this paper we will focus on  phenomenology of $\alpha$-attractors in hybrid inflation. Therefore in the main part of the paper for simplicity we will consider models describing  real scalar fields, but  our results can be also formulated   in terms of complex fields, in context of supergravity, see Appendix A.

While the plateau shape of the potential is a generic property of all $\alpha$-attractors, the  approach to the plateau can be slightly different.
 
In exponential $\a$-attractors \cite{Kallosh:2013yoa}, where the field approaches the plateau exponentially fast,  in the large $N$ limit, where $N$ is the number of e-foldings,  one has 
\be \label{exp}
V = V_{0} (1-e^{-\vp/\mu}+\dots)  \, , \qquad \mu = \sqrt{3\a\over 2}  \, , \qquad n_s\approx 1-{2\over N}\, , \qquad r\approx {12 \a\over N^2}  \ .
\ee 
for $\mu \lesssim O(1)$. For example, for $N =55$ 
\be
n_s\approx 1-{2\over 55} \approx 0.963 \ .
\ee
Predictions of the simplest models of this class can completely cover the left part of the $n_{s}-r$ area favored by the latest Planck/BICEP/Keck data \cite{Kallosh:2021mnu}, nearly independently of the choice of the original inflaton potential.

For the family of polynomial $\a$-attractors \cite{Kallosh:2022feu}, where   the potential approaches a plateau as inverse powers of the inflaton field, one has
\be \label{pol}
 V \sim V_{0}\large(1 -{\mu^{k}\over \vp^{k}}+\dots \large)\, , \qquad n_{s} = 1-{2\over N}{k+1\over k+2}\, , \qquad r\approx { 8k^2 \mu^{2k\over k+2}\over 
[(k(k+2))N]^{2k+ 2\over k+2}} \ .
 \ee
Here $k$ can take any positive value. For example, in $k=2$ case
\be \label{pol2}
 V \sim V_{0}\large(1 -{\mu^{2}\over \vp^{2}} + \dots \large)\, , \qquad n_{s} \approx 1-{3\over 2 N} \, , \qquad r \approx {\sqrt 2 \, \mu\over N^{3/2}}
 \ee
For $N=N_{\rm total} =55$ we have 
\be
n_s\approx 1-{3\over 110}\approx 0.973 \ .
\ee 
By taking smaller $k$, one can increase the value of $n_{s}$ in this scenario from $1-{2\over N}$ to  $1-{1\over N}$.
As a result, predictions of the simplest models of  exponential and polynomial attractors   completely cover the  $n_{s}-r$ area favored by the latest Planck/BICEP/Keck data, see Fig. 3 of \cite{Kallosh:2022feu}.

Thus it would seem that a rather simple set of models of this type can describe any set of data which any future observations may bring.  However, there are still some issues which one may try to address. 

1) One may wonder whether it is possible to increase $n_{s}$ to cover the right part of the $n_{s}-r$ area favored by the latest Planck/BICEP/Keck data within the  more familiar class of exponential $\alpha$-attractors \rf{exp}. 

2) There are ongoing efforts to solve the $H_{0}$ and $S_{8}$ problems by modifying the standard $\Lambda$CDM model  \cite{Riess:2021jrx,Abdalla:2022yfr}. Some of these efforts require  a significant re-interpretation of the available data, resulting in much higher values of $n_{s}$, all the way up to the  Harris-Zeldovich value $n_{s} = 1$, see \cite{Jiang:2022uyg,Smith:2022hwi} and references therein. Thus one may wonder whether one may find  some versions of $\alpha$-attractors which would be compatible with such values of $n_{s}$.  

3) In models of $\alpha$-attractors inspired by string theory and M-theory, one may encounter  many interacting scalar fields, each of which may have inflaton potentials with different values of $\alpha$  \cite{Ferrara:2016fwe,Kallosh:2017ced,Kallosh:2017wnt,Achucarro:2017ing,Yamada:2018nsk,Linde:2018hmx,Gunaydin:2020ric,Kallosh:2021fvz,Kallosh:2021vcf,Kallosh:2022vha}. 
Therefore it is important to explore multi-field $\alpha$-attractors. In the simplest cases, one may have several different stages of inflation, but in many models the last  $N  \sim 50$ - 60   e-foldings of inflation are described by  a single stage of inflaton, with the predictions described above. 

However, this is not always the case. For example, suppose that there is a short secondary   stage of inflation  describing $\Delta N$ e-foldings after the $\alpha$-attractor stage. In this case, we must carefully distinguish between the total number of e-foldings $N_{e}  \sim 50$ - 60  responsible for the observable structure of the universe, and its part $N$ related to inflation in the $\alpha$-attractor regime:
\be
N = N_{e}  - \Delta N \ .
\ee
The observational predictions of $\alpha$-attractors are still described by \rf{exp}, \rf{pol2}, but the value of  $N = N_{e}- \Delta N$ becomes smaller than $N_{e}  \sim 50$ - 60 \cite{Christodoulidis:2018qdw,Linde:2018hmx}. This may significantly decrease the value of $n_{s}$, which may contradict the observational data unless the second stage of inflation is very short.

This issue is less important for polynomial attractors \rf{pol} because they predict higher values of $n_{s}$. That is why some of the popular models of large PBH formation  \cite{Braglia:2020eai} can be formulated in the context of the KKLTI polynomial $\alpha$-attractors  \cite{Kallosh:2022vha}, whereas similar models based on exponential $\alpha$-attractors tend to predict very small PBHs \cite{Iacconi:2021ltm}. It would be interesting to see whether one may overcome these limitations   and find a way to increase $n_{s}$, if required.

In this paper we will show how one can significantly increase $n_{s}$ in two-field inflationary models. The main mechanism which we are going to discuss is rather general. As an example, we will study the original version of the hybrid inflation scenario \cite{Linde:1991km,Linde:1993cn}, and then explore its $\alpha$-attractor implementation.  In these models, the potential of the inflaton field $\vp$ is  uplifted by the potential of the second field $\chi$, but this uplifting ends   due to a tachyonic instability with respect to the field $\chi$, which happen when the field $\vp$ becomes smaller than its critical value $\vp_{c}$. This instability typically leads to a nearly instant end of inflation and rapid reheating, but it may also occur slowly, in a secondary inflationary stage. 

 We will confirm that the main attractor predictions \rf{exp}, \rf{pol2} remain true in these models in the large $N$ limit. However, we will show that in some models the large $N$ limit is achieved only for $N > 60$, and  for  $N \lesssim 60$ one may have an intermediate asymptotic regime with $n_{s}$ that can be  greater than the attractor values  \rf{exp}, \rf{pol2}. 
In particular, for any fixed $N$ (e.g. for $N \sim 50$), in the  large uplift limit, or in the limit of large value of $\vp_{c}$,  we find another attractor prediction, the Harrison-Zeldovich spectrum with $ n_s \rightarrow  1$.

\section{\boldmath{Single field $\alpha$-attractors}}\label{sattr}
 
We will begin with describing single field $\alpha$-attractors.
The simplest example is given by the theory
 \be
{ {\cal L} \over \sqrt{-g}} =  {R\over 2}  -  {(\partial_{\mu} \phi)^2\over 2\bigl(1-{\phi^{2}\over 6\alpha}\bigr)^{2}} - V(\phi)   \,  .
\label{cosmoA}\ee
Here $\phi(x)$ is the scalar field, the inflaton.  In the limit $\alpha \to \infty$ the kinetic term becomes the standard canonical term $-  {(\partial_{\mu} \phi)^2\over 2}$.  The new kinetic term has a singularity at $|\phi| = \sqrt{6\alpha} $. However, one can get rid of the singularity and recover the canonical normalization  by solving the equation ${\partial \phi\over 1-{\phi^{2}\over 6\alpha}} = \partial\vp$, which yields $
\phi = \sqrt {6 \alpha}\, \tanh{\varphi\over\sqrt {6 \alpha}}$.
The full theory, in terms of the canonical variables, becomes a theory with a plateau potential
 \be
{ {\cal L} \over \sqrt{-g}} =  {R\over 2}  -  {(\partial_{\mu}\varphi)^{2} \over 2}  - V\big(\sqrt {6 \alpha}\, \tanh{\varphi\over\sqrt {6 \alpha}}\big)   \,  .
\label{cosmoqq}\ee
We called such models T-models due  to their dependence on the $\tanh{\varphi\over\sqrt {6 \alpha}}$.  Asymptotic behavior of the potential at  large $\varphi>0$ is given by
\be\label{plateau}
V(\vp) = V_{0} - 2  \sqrt{6\alpha}\,V'_{0} \ e^{-\sqrt{2\over 3\alpha} \varphi } \ .
\ee
Here $V_0 = V(\phi)|_{\phi =  \sqrt {6 \alpha}}$ is the height of the plateau potential, and $V'_{0} = \partial_{\phi}V |_{\phi = \sqrt {6 \alpha}}$. The coefficient $2  \sqrt{6\alpha}\,V'_{0}$ in front of the exponent  can be absorbed into a redefinition (shift) of the field $\varphi$. Therefore  inflationary predictions of this theory in the regime with $e^{-\sqrt{2\over 3\alpha} \varphi } \ll 1$ are determined only by two parameters, $V_{0}$ and $\alpha$, i.e. they do not depend on many other features of the potential $V(\phi)$. That is why they are called attractors.
 
 At large $N$,  predictions of these models for $A_{s}$, $n_{s}$ and $r$ coincide in the small $\alpha$ limit, nearly independently of the detailed choice of the potential $V(\phi)$:
\be
\label{pred}
 A_{s} = {V_{0}\, N^{2}\over 18 \pi^{2 }\alpha} \ , \qquad n_{s} = 1-{2\over N} \ , \qquad r = {12\alpha\over N^{2}_{e}} \ .
\ee 
These models are compatible with the presently available observational data for sufficiently small $\alpha$.

Importantly, these results depend on the height of the inflationary plateau, which is given by $V_0 = V(\phi)|_{\phi =  \sqrt {6 \alpha}}$, but they do not depend on many other details of behavior of the  potential $V(\phi)$ in \rf{cosmoA}. This explains, in particular, stability of the predictions of these models with respect to quantum corrections \cite{Kallosh:2016gqp}.

The amplitude of inflationary perturbations  in these models matches the Planck normalization $A_{s} \approx 2.01 \times 10^{{-9}}$ for  $ {V_{0}\over  \alpha} \sim 10^{{-10}}$, $N = 60$, or for $ {V_{0}\over  \alpha} \sim 1.5 \times 10^{{-10}}$, $N = 50$.  For the simplest  model $V = {m^{2}\over 2} \phi^{2}$  one finds
\be\label{T}
V =  3m^{2 }\alpha \tanh^{2}{\varphi\over\sqrt {6 \alpha}} \ .
\ee
This simplest model  is shown by the prominent vertical yellow band in Fig. 8 of the paper on inflation in the Planck2018 data release  \cite{Planck:2018jri}.  In this model,  the condition $ {V_{0}\over  \alpha}  = 3 m^{2}=\sim 10^{{-10}}$ reads $ m  \sim   0.6 \times10^{{-5}}$. The small magnitude of this parameter  accounts for the small amplitude of perturbations $A_{s} \approx 2.01 \times 10^{{-9}}$. No other parameters are required to describe all presently available inflation-related data in this model. If the inflationary gravitational waves are discovered, their amplitude can be accounted for by the  choice of the parameter $\alpha$ in \rf{pred}.

The results described above are valid under assumptions that the potential $V(\phi)$ and its derivatives are non-singular at the boundary $|\phi| = \sqrt{6\alpha} $. If one keeps the requirement that the potential $V(\phi)$ is non-singular, but allows its derivatives to be singular, the potential $V(\vp)$ remains a plateau potential in canonical variables, but it may become a polynomial attractor, with properties and predictions described in \rf{pol}, \rf{pol2}  \cite{Kallosh:2022feu}.

One should note also,  that these results  rely on a hidden assumption that inflation occurs in the single field regime with a potential \rf{exp} or  \rf{pol}, and ends when the slow-roll conditions  are no longer satisfied. This assumption is natural indeed, but one can find, or engineer, some models where it may be violated.

As we already mentioned in the previous section,  the  simplest possibility  to do it is to  arrange for a second stage of inflation with duration  $\Delta N$.  This modification decreases $n_{s}$. For exponential $\alpha$-attractors \rf{exp} this decrease is not particularly desirable.

However, there is yet another possibility, which  may allow many interesting variations of the main theme. One may consider multi-field models, where the single-field inflation regime ends prematurely because of the instability of the inflationary trajectory, or because of its sharp turn. 

The simplest well-known example is provided by hybrid inflation  \cite{Linde:1991km,Linde:1993cn}. In this scenario,  inflation driven by the field $\phi$ is terminated because of the tachyonic waterfall instability with spontaneous generation of the second field $\sigma$. This mechanism involves two ingredients, each of  which allow to control (increase) $n_{s}$. First of all, this scenario involves uplift of an inflationary potential by some potential depending on $\sigma$. This uplift disappears after the waterfall instability, but during inflation with $\phi > \phi_{c}$ the uplift increases $V$ while keeping $V'$ intact. This decreases slow-roll parameters and increases $n_{s}$ for $\phi > \phi_{c}$. Secondly, one can control the value of  $\phi_{c}$ by a proper choice of parameters. As a result, one can also control the value of the field $\phi_{N}$ corresponding to $N$ e-foldings prior to termination of inflation. This provides an additional tool to control $n_{s}$.

In this paper we will consider hybrid models of $\alpha$-attractors and explain how both of these mechanisms  affect inflationary predictions for $n_{s}$ and $r$. To avoid misunderstandings, we should emphasize  that hybrid $\alpha$-attractors are  more complicated than the  single-field $\alpha$-attractors. However, realistic inflationary models often involve more than one scalar field. As we will see, investigation of their $\alpha$-attractor versions can be quite instructive.

\section{Hybrid inflation}\label{oldhybrid}

\subsection{Original hybrid inflation model}

\begin{figure}[H]
\centering
\includegraphics[scale=0.4]{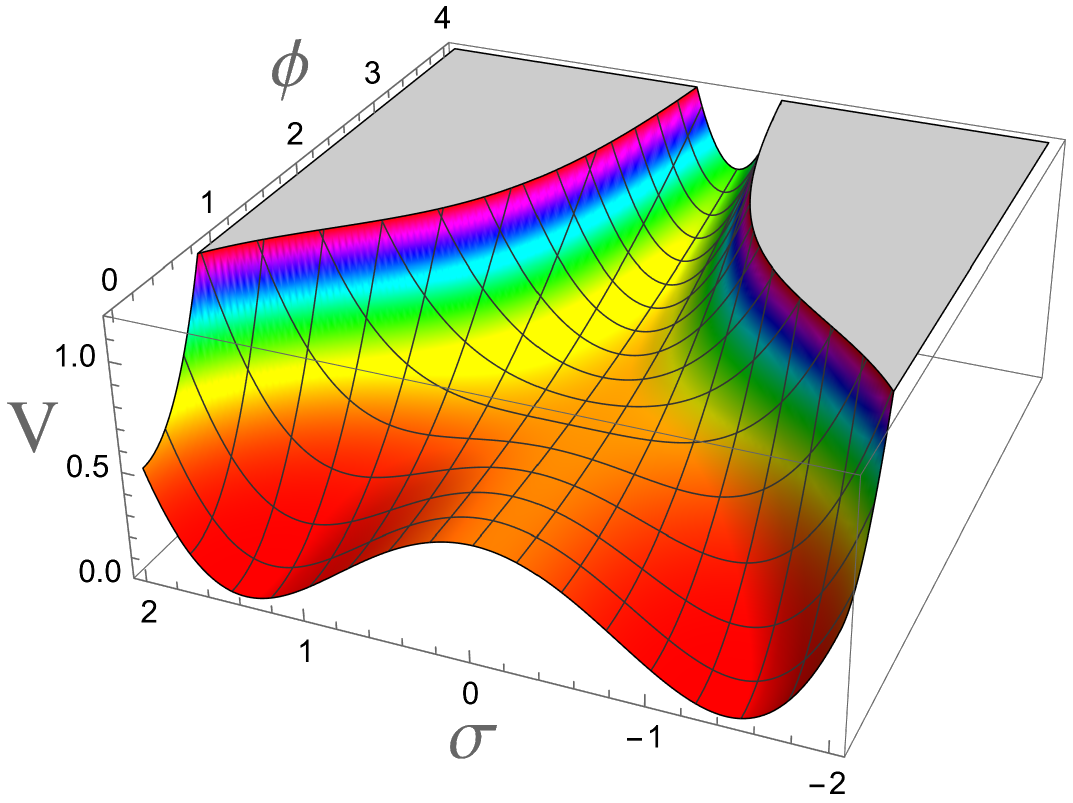}
\caption{\footnotesize  Hybrid inflation potential \rf{hybrid} for $m = 0.2, M = 1, \lambda = 0.5, g = 0.8$.}
\label{h1}
\end{figure}

Let us first consider the simplest hybrid inflation model \cite{Linde:1991km,Linde:1993cn}. The
effective potential of this model is given by
\begin{equation}\label{hybrid}
V(\sigma,\phi) =  {1\over 4\lambda}(M^2-\lambda\sigma^2)^2
+ {m^2\over
2}\phi^2 + {g^2\over 2}\phi^2\sigma^2\ .
\end{equation}
To illustrate the main features of this potential, we show it in Fig. \ref{h1}.

The effective
mass squared of the field $\sigma$ at $\sigma = 0$ is equal to  
\be
V_{ \sigma,\sigma}(\sigma = 0) = -M^{2}+g^{2}\phi^{2} \ .
\ee 
For $\phi > \phi_c = M/g$ the only minimum of the effective
potential
$V(\sigma,\phi)$ with respect to $\sigma$  is at $\sigma = 0$. The curvature of the effective
potential in the $\sigma$-direction is much greater than in the 
$\phi$-direction. Thus we expect that at the first stages of
expansion of the Universe the field $\sigma$ rolled down to $\sigma =
0$, whereas the field $\phi$ could remain large for a much longer
time.    

The potential at $\sigma = 0$ can be written as 
\be
V(\sigma= 0,\phi) = V_{\rm up} +  {m^2\over 2}\phi^2 \ ,
\ee
where the uplifting potential is 
\be 
V_{\rm up} = {M^4\over 4\lambda}  \ .
\ee

At the moment when the inflaton field $\phi$ becomes smaller than  $\phi_c = M/g$, the phase transition
with the symmetry breaking occurs. For a proper choice of parameters, this phase transition  occurs very fast, and inflation abruptly ends \cite{Linde:1991km,Linde:1993cn}. However, there are some situations where inflation may continue for a while in the process of spontaneous symmetry breaking, which may lead to  production of primordial black holes (PBHs) \cite{Garcia-Bellido:1996mdl}.

Unfortunately, these models are disfavored by the data in most of its parameter space: at ${m^2\over
2}\phi^2\gtrsim V_{\rm up}$ the  tensor-to-scalar ratio is too  high, whereas at ${m^2\over
2}\phi^2\ll V_{\rm up}$ the spectral index $n_{s}$ is too high:  $n_s >1$  \cite{Planck:2013jfk}.

 Once we switch to $\alpha$-attractor version of hybrid inflation, the first of these problems disappears. As we will show later, the second problem may also disappear: in the large $N$ limit these models lead to the standard $\alpha$-attractor predictions \rf{exp}, \rf{pol}.  The issue we need to carefully examine is whether $N\sim 60$ is large enough to be described by the large $N$ limit.  
 
Before we switch to $\alpha$-attractors we should mention a property of such models, which may be either a problem or an advantage. As one can see from Fig. \ref{h1}, at the $\phi < \phi_c$  the field $\sigma$ may fall into one of the two minima of the potential, at $\sigma = \pm M/\sqrt \lambda$ This may divide the universe into many domains with $\sigma = \pm M/\sqrt \lambda$ separated by domain walls. Unless $V_{\rm up}$ is extremely small, this leads to unacceptable cosmological consequences.  
 
 The simplest way to avoid this problem  is to study models where the field $\sigma$ is a complex field. Then, instead of domain walls, one has cosmic strings \cite{Linde:1993cn}. If $M/\sqrt \lambda$ is not too large, these strings may have interesting cosmological implications. On the other hand, in the models with large magnitude of symmetry breaking, one may want to avoid productions of  topological defects. The simplest possibility is to add a tiny linear term $c  \sigma$ to the potential \rf{hybrid}. If this term is very  small,  it leads only to a minor tilt of the potential towards one of the directions, which may be sufficient  to eliminate the production of the topological defects, while leaving other  predictions of the scenario intact. Other ways to avoid production of topological defects can be found in \cite{Lazarides:1995vr,Jeannerot:2000sv}. In the next section and in the Appendix we will describe two novel mechanisms which can suppress production of the topological defects in the context of $\alpha$-attractors.

\subsection{Hybrid   $\alpha$-attractors}

\begin{figure}[H]
\centering
\includegraphics[scale=0.4]{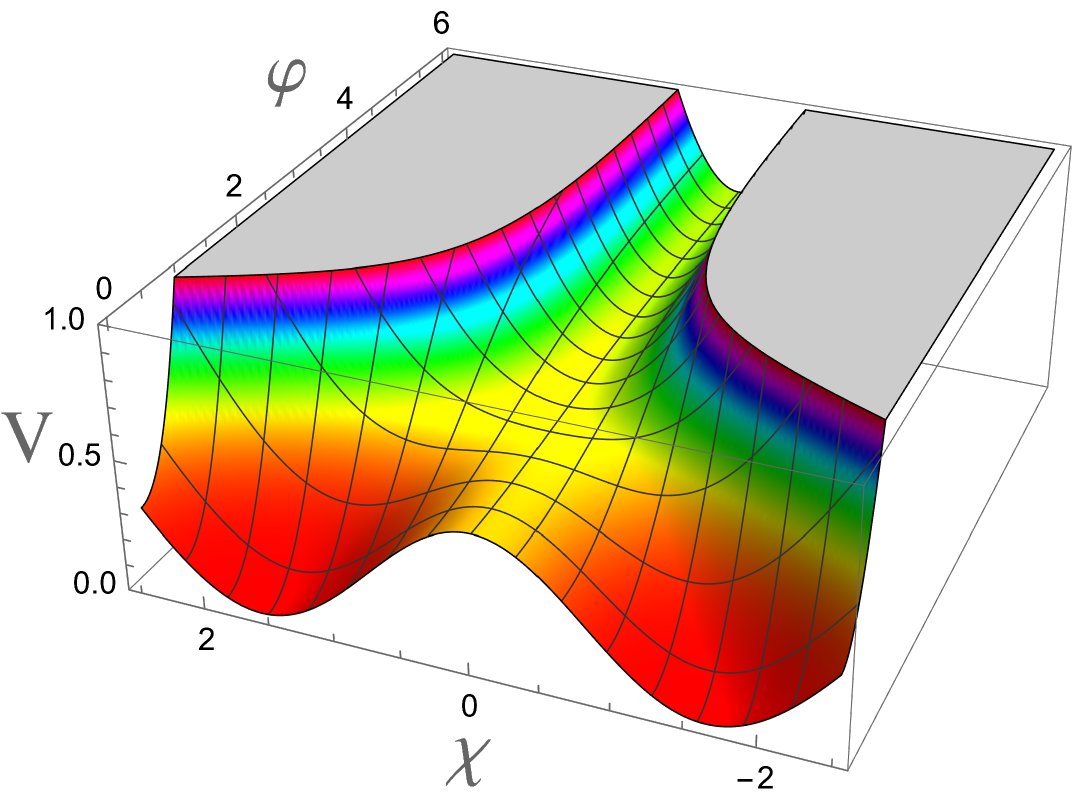}
\caption{\footnotesize  Hybrid inflation potential for the model \rf{hybridab} with $m = 0.2, M = 1, \lambda = 0.5, g = 0.8, \alpha = 1, \beta = 1$. It looks very similar to the original potential shown in Fig. \ref{h1}, but the potential along the valley $\chi = 0$ is much more flat, see Fig. \ref{h21d}. }
\label{h2}
\end{figure}

Here we will explore what may happen if we generalize the hybrid inflation model \rf{hybrid}
by embedding it in the context of exponential $\alpha$-attractors.  We will discuss polynomial attractors \cite{Kallosh:2022feu} in section \ref{polS}.
 \be 
{ {\cal L} \over \sqrt{-g}} =  {R\over 2}  -  {(\partial_{\mu} \phi)^2\over 2\bigl(1-{\phi^{2}\over 6\alpha}\bigr)^{2}} -  {(\partial_{\mu} \sigma)^2\over 2\bigl(1-{\sigma^{2}\over 6\beta}\bigr)^{2}} - V(\sigma,\phi)   \,  .
\label{cosmoAA}\ee 
Upon  a transformation to canonical variables $\vp$ and $\chi$, the hybrid inflation potential becomes
\bea\label{hybridab}
V(\chi,\vp) &=&  {1\over 4\lambda}\large(M^2-6\beta\lambda \tanh^{2}{\chi\over\sqrt {6 \beta}}\large)^2
+3m^{2 }\alpha \tanh^{2}{\varphi\over\sqrt {6 \alpha}}\nonumber \\ &+& 18 g^2\alpha \beta\,  \tanh^{2}{\varphi\over\sqrt {6 \alpha}} \,  \tanh^{2}{\chi\over\sqrt {6 \beta}}.
\eea
The shape of this potential for some particular values of parameters is shown in Fig. \ref{h2}.

\begin{figure}[H]
\centering
\includegraphics[scale=0.5]{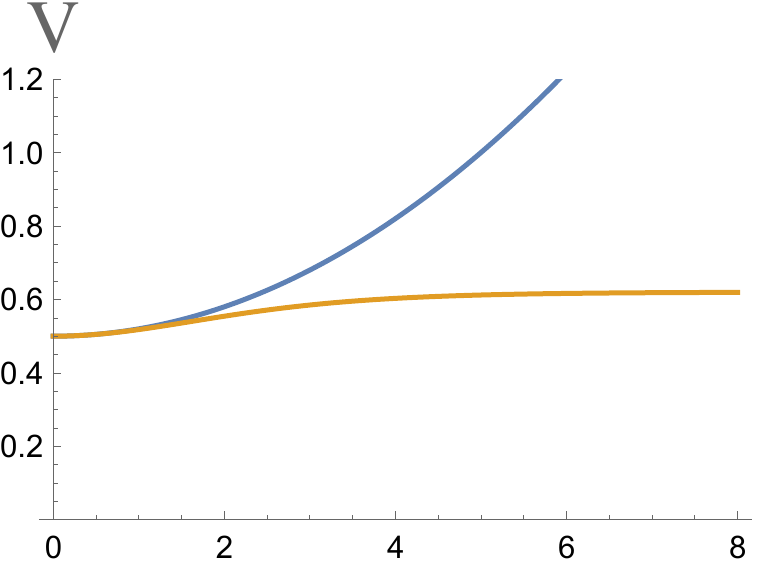}
\caption{\footnotesize  The blue line shows the original potential \rf{hybrid}  along the flat direction $\phi$ for $\sigma = 0$ and $\phi< 5$ The brown line shows the potential of the $\alpha$-attractor  \rf{hybridab} for $\alpha = 1$ along the flat direction $\varphi$ for $\chi = 0$ and $\vp <5$. Note that the $\alpha$-attractor potential  is much more flat, because the full potential $V(\vp)$ is produced by the horizontal stretching of the part of the potential $V(\phi)$ with $\phi < \sqrt{6 \alpha}$.   }
\label{h21d}  
\end{figure}

In Fig. \ref{h21d} we show by the blue line the original potential \rf{hybrid}  along the flat direction $\phi$ for $\sigma = 0$, and we also show by the brown line the potential of the $\alpha$-attractor  \rf{hybridab} for $\alpha = 1$ along the flat direction $\varphi$ for $\chi = 0$.  It illustrates the flattening of the inflaton potential for $\alpha$-attractors.

The curvature of the potential in the $\chi$ direction at $\chi = 0$ coincides  with the curvature with respect to $\sigma $ at $\sigma = 0$:
\be
 V_{ \chi,\chi}(\chi = 0)= V_{ \sigma,\sigma}(\sigma = 0) = -M^{2}+g^{2}\phi^{2}=  -M^{2}+6\alpha\, g^{2} \tanh^{2}{\varphi\over\sqrt {6 \alpha}}.
\ee 
For $\phi > \phi_{c} = M/g$, this curvature is positive, and the inflationary trajectory with $\chi = 0$ remains stable until field $\phi$ rolls below the critical point
\be\label{crit}
\phi_{c} = \sqrt {6\alpha}  \tanh {\varphi_{c}\over\sqrt {6 \alpha}} = M/g \ .
\ee

 If the last 60 e-foldings of inflation occur when $|\phi| \ll \sqrt{6\alpha}$, $|\sigma| \ll \sqrt{6\beta}$, then most cosmological consequences of this  model will coincide with those of the original version of hybrid inflation \cite{Linde:1991km,Linde:1993cn}.

Notice that in the limit when $|\phi| \ll \sqrt{6\alpha}$, $|\sigma| \ll \sqrt{6\beta}$, the kinetic terms in eq. \rf{cosmoAA} become canonical, and therefore the shape of the potential reduces  to the one in the original version of hybrid inflation.
In particular, in the large $\alpha$ limit inflation ends at $\phi_{c} \approx  \vp_{c} = M/g$. In this paper we will be interested in the opposite possibility, when the last 60 e-foldings occur in the $\alpha$-attractor  regime where $\varphi_{c} \gg \sqrt {6 \alpha}$.

\begin{figure}[t!]
\centering
\includegraphics[scale=0.4]{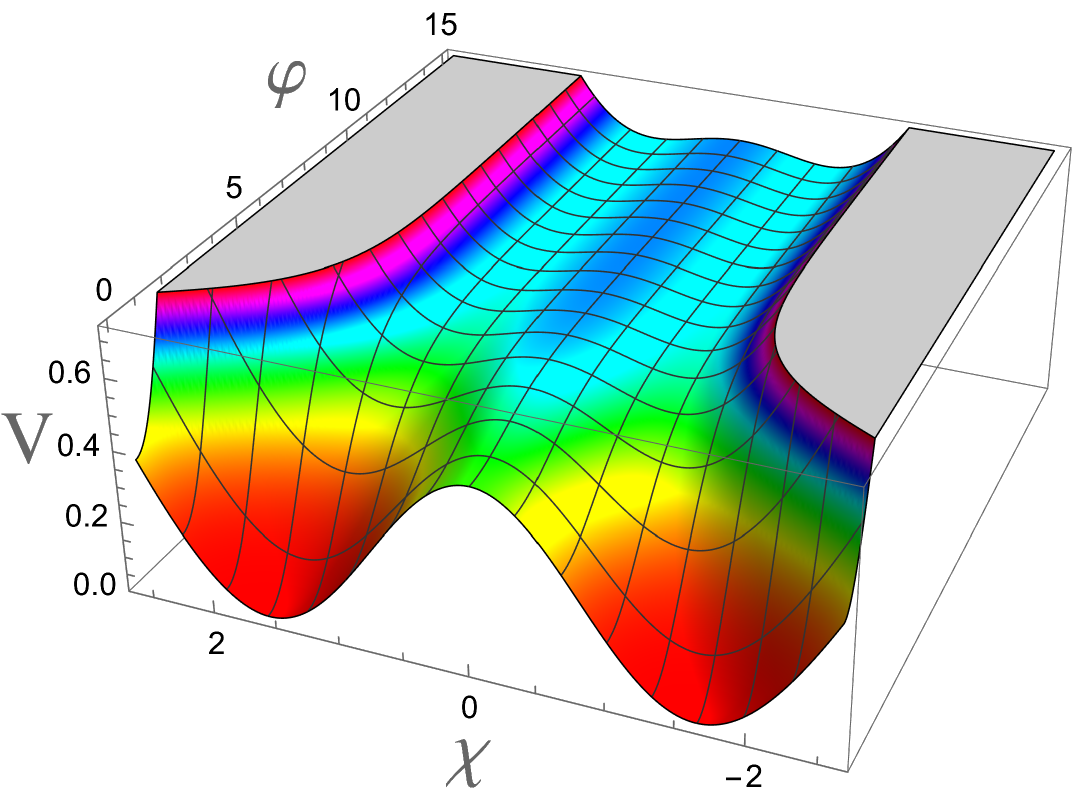} \includegraphics[scale=0.4]{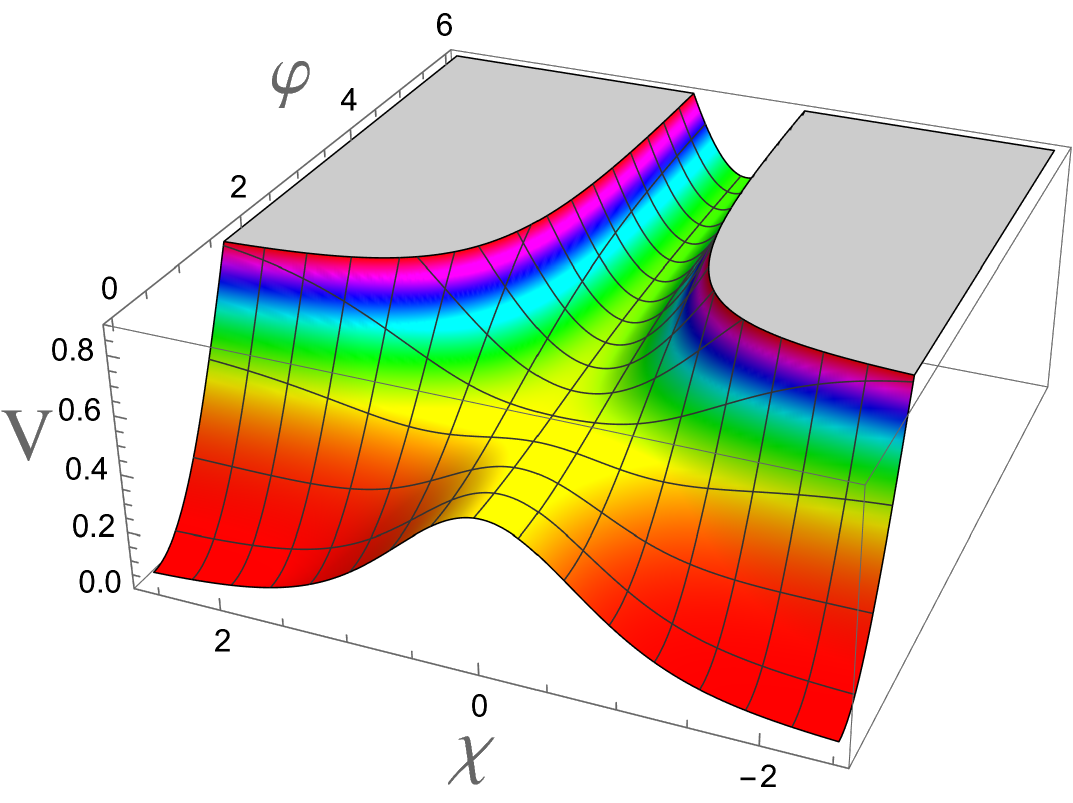}
\caption{\footnotesize   Left panel shows  potential \rf{hybridab} for $m = 0.2, M = 1, \lambda = 0.5, g = 0.35, \alpha = 1, \beta = 1$.  Right panel shows  potential \rf{hybridab} for $m = 0.2, M = 1, \lambda = 0.5, g = 0.8, \alpha = 1, \beta = 1/4$.}
\label{h3}
\end{figure}

 One should note also that the standard scenario with the waterfall phase transition shown in Fig. \ref{h2}   occurs only if $\phi_{c} = M/g  < \sqrt{6 \alpha}$. In the opposite case $\phi_{c} = M/g  > \sqrt{6 \alpha}$  the field $\chi$ does not vanish at any values of $\vp$, because all  values of $\vp$ correspond to $\phi  < \sqrt{6 \alpha}$. The amplitude of spontaneous symmetry breaking grows during inflation starting from $\chi^{2} = {M^{2} - 6\alpha \over \lambda}$ at $\vp \to \infty$, and gradually approaching its maximal value $\chi^{2} = {M^{2}\over \lambda}$ at $\phi = 0$. Since the symmetry breaking with respect to the sign of the field $\chi$ is present from the very beginning of inflation, see the left panel of Fig. \ref{h3}, topological defects do not form in this scenario. Thus it does  not suffer from any problems with topological defects which may appear in the scenario shown in Figs. \ref{h1}, \ref{h2}, see   the previous section.

To illustrate what happens for $M/g  > \sqrt{6 \alpha}$, we plot in the left panel of Fig. \ref{h3} the potential \rf{hybridab} for the same values of parameters as in Fig. \ref{h2}. The only parameter we change is $g$, which we take smaller, $g = 0.35$. 
 
This is not the last of the surprises which may await us after introducing hybrid $\alpha$-attractors, see the right panel in Fig. \ref{h3}, where we plot the same potential for the same parameters as in Fig. \ref{h2}, but for a smaller value of $\beta$.
 As we see, in this case the position of the minimum of the potential with respect to $\chi$ disappears, and we end  up with the potential describing the $\alpha$-attractor generalization \cite{Dimopoulos:2017zvq,Akrami:2017cir,Braglia:2020bym} of the quintessential inflation  \cite{Peebles:1998qn,Felder:1999pv}. This happens because for sufficiently small $\beta$ the position of the minimum of the potential with respect to $\sigma$ moves outside the boundary of the moduli space at $\sigma = \sqrt{6\beta}$.

It is not our goal to describe all of these interesting possibilities in this paper. In what follows we will study the more traditional regime described by Fig. \ref{h2}. In this regime, the initial stages of inflation occur at $\chi = 0$, until the field reaches a critical point  $\vp_{c}$. After that, the tachyonic instability with respect to the field $\chi$ terminates the stage of inflation at $\chi = 0$. Depending on the parameters of the model, this may lead either to an abrupt end of inflation,  or to a beginning of a short additional period of inflation.  We will focus on the first of these two possible outcomes, and calculate inflationary parameters $A_{s}$, $n_{s}$ and $r$ assuming that inflation ends at the moment when the field $\vp$ reaches $\vp_{c}$ \rf{crit}.   

Inflationary potential at $\chi = 0$ is given by
\be 
V(\vp)=    {M^{4}\over 4\lambda}+  3m^{2 }\alpha \tanh^{2}{\varphi\over\sqrt {6 \alpha}}  \ .
\ee
Using equation \rf{plateau}, one can represent this potential during inflation at $\vp \gg \sqrt {6 \alpha}$ in this model as
\be
V = V_{\rm up} +   V_{0} \, (1-4  \ e^{-\sqrt{2\over 3\alpha} \varphi } +...) \ ,
\ee
where $ V_{\rm up} = {M^{4}\over 4\lambda}$ is the value of the uplifting potential ${1\over 4\lambda}(M^2-\lambda\sigma^2)^2$ at $\sigma = 0$, and $V_{0}= 3m^{2 }\alpha$ is the value of the $\alpha$-attractor potential $3m^{2 }\alpha \tanh^{2}{\varphi\over\sqrt {6 \alpha}}$ at its plateau.

Let us first consider the regime  $V_{up}\gg V_{0}$, i.e 
\be
M^{4} \gg  12\alpha\lambda  m^{2} \ .
\ee
The Hubble constant in this case is 
\be
H^{2}  = {M^{4}\over 12  \lambda} \ .
\ee
Thus $ M^{2} \gg H^{2}$ for 
\be 
M^{2} \ll  12  \lambda  \ .
\ee

If $M^{2} \ll  12  \lambda$, then shortly after  the field $\phi$ moves  below the critical value $\phi_{c} = M/g$, the effective mass squared of the field $\chi$ becomes negative. Once its absolute value becomes  greater  than $H^{2}$,  the tachyonic instability of the field $\chi$ develops, which leads to an  abrupt termination of inflation at $\phi \approx \phi_{c}$, as in the standard version of the hybrid inflation scenario  \cite{Linde:1991km,Linde:1993cn}.
  
\section{{\boldmath Inflationary predictions  of hybrid $\alpha$-attractors}}\label{predictions}

In our investigation of perturbations in the hybrid inflation, we will try to be as model-independent as possible. The results to be obtained in this section will be applicable not only to  hybrid inflation, but to any $\alpha$-attractor potentials uplifted by an additional term similar to the first term in \rf{hybrid}. We will also assume that the single-field regime may end  not because of the violation of the slow-roll conditions, but because some kind of instability terminating the original stage of inflation in a vicinity of a critical field $\phi_{c}$, as in the hybrid inflation scenario.

The general $\alpha$-attractor potential \rf{plateau} at large $\vp$ can be represented as
\be\label{plateau1a}
V(s) = V_{0}\large(1 -  e^{-\sqrt{2\over 3\alpha} s}+...\large) \ ,
\ee
where $s$ is given by
\be\label{sf}
s = \vp - \sqrt{3\alpha\over 2} \ln \Bigl(2  \sqrt{6\alpha}\,{V'_{0}\over V_{0}}\Bigr)  \ ,
\ee
and  $V'_{0} = \partial_{\phi}V |_{\phi = \sqrt {6 \alpha}}$ at the boundary $\phi = \sqrt{6\alpha}$, as in \rf{plateau}. To give a particular example, in the simplest T-model \rf{T}   one has 
\be
s = \vp - \sqrt{3\alpha\over 2} \ln 4  \approx \vp -1.7 \sqrt{\alpha} \ .
\ee
Thus for $\alpha \lesssim 1$ one has $\vp = s +O(1)$.

 Now we will uplift this potential by adding to it a constant $V_{up}$. In the hybrid inflation model \rf{hybrid} one has $V_{\rm up} = {M^{4}\over 4\lambda}$. The full potential becomes
\be\label{plateau1}
V(s) =V_{\rm up}+  V_{0}(1 -  e^{-\gamma s}) \ ,
\ee
where $\gamma$ is related to the \K curvature 
\be
\gamma = \sqrt{2\over 3\alpha}\, , \qquad \gamma^2= {2\over 3\alpha}  \ .
\ee
This form correctly describes the potential for  
\be
e^{- \gamma s}\ll 1 \ .
\ee 
 We consider a stage of $N \gg 1$ e-foldings of inflation which begins at $s_{N}$ and ends at $s_{c}$. Inflation may continue when the field reaches $s_{c}$, or it may end abruptly if the inflationary trajectory  changes at $s_{c}$ because of the waterfall instability at   in hybrid inflation.

Equation describing evolution of $s$ in the slow-roll regime is
\be
{ds \over dN} =  {{dV\over ds}\over V(s)} = { V_{0} \gamma \ e^{- \gamma\,  s }\over V_{\rm up}+  V_{0}(1 -  \ e^{- \gamma\,  s })} \ .
\ee
We are interested in the regime $e^{- \gamma\,  s } \ll 1$. In that case one can ignore the exponent in the denominator and find  a solution of this equation:

\be\label{solution}
e^{\gamma\, s_N} =   { \gamma^{2} V_{0} N\over  V_{\rm up}+V_{0} } + e^{\gamma\, s_c} \ .
\ee
where $s_N$ is the value of the field $s$ at $N$ e-foldings before the end of this stage of inflation before it reaches $s_c$, i. e. $s_N= s_c$ at $N=0$.

The standard expression for $n_{s}$ is 
\be
n_{s} = 1-3\left({V'\over V}\right)^{2}+ 2{V''\over V} \approx 1- {  3V_{0}^{2} \gamma^{2}\ e^{- 2\gamma\,  s_N }\over  (V_{\rm up}+ V_{0})^{2}}- { 2 V_{0} \gamma^{2}   e^{- \gamma\,  s_N }\over  V_{\rm up}+V_{0} } \approx 1- { 2 V_{0} \gamma^{2}   e^{- \gamma\,  s_N }\over  V_{\rm up}+V_{0} } \ .
 \ee
Here  the derivatives are taken with respect to $s$. Using equation \rf{solution}, we find
\be
n_s=1- \frac{2 V_{0} \gamma^2}{V_{0}\, \gamma^2 N +(V_{\rm up}+ V_{0})  e^{\gamma s_c}} \ .
\ee
In the large $N$ limit we always have the standard universal $\alpha$-attractor prediction, independently of all other parameters of the model,
\be\label{nsnormal}
n_{s} =1-  {2  \over  N}  \ .
\ee
However,  the range accessible to observations is limited, $N \lesssim 50 - 60$. 
For 
\be\label{gg}
  e^{\gamma s_c} \gg  { \gamma^{2} V_{0} N\over  V_{\rm up}+V_{0} }  \ ,
 \ee 
 one has, in accordance with \rf{solution},
\be\label{equality}
e^{\gamma s_N} \approx e^{\gamma s_c}  ,
\ee
and instead of the large $N$ limit, one has a different limiting case,
\be
1-n_s=  { 2 V_{0} \gamma^{2}   e^{- \gamma\,  s_N }\over  V_{\rm up}+V_{0} } \approx  { 2 V_{0} \gamma^{2}   e^{- \gamma\,  s_c }\over  V_{\rm up}+V_{0} } \ll {2\over N}\ ,
\ee
where the last inequality follows from \rf{gg}. Thus in the large $V_{\rm up}$ limit (for large ratio $V_{\rm up}/V_{0}$), or in the large $s_{c}$ limit (for $\gamma s_c \gg 1$), when inequality \rf{gg} is satisfied, we have $n_{s} \to 1$, i.e. the Harrison-Zeldovich spectrum.

Interpolating between these two limiting cases by changing $V_{\rm up}/V_{0}$, or by changing $s_{c}$, one can find any value of $n_{s}$ in the  range 
\be
1-  {2  \over  N} \ \lesssim \ n_{s} \ \lesssim \ 1.
\ee
In particular, for 
\be\label{inter}
 V_{\rm up}+ V_{0}  = V_{0}\, \gamma^2 N  e^{-\gamma s_c}
 \ee
 we have 
\be\label{half}
n_{s} = 1 - {1\over N}  \ .
\ee
Let us consider the  implications for the amplitude of perturbations $A_{s}$ and for $r$.
 \be
 A_{s} = {V^{3}\over 12 \pi^{2}(V')^{2}} \approx {  (V_{\rm up}+  V_{0})^{3}\ e^{ 2\gamma\,  s_N }\over 12 \pi^{2} V_{0}^{2} \gamma^{2} }       \ .
 \ee
 In the large $N$ limit one finds
\be\label{newampl}
 A_{s} =  {(V_{\rm up}+  V_{0})   N^{2}\over 18 \alpha \pi^{2}    \   } \ ,
 \ee
 Meanwhile for $ V_{\rm up}+ V_{0}    \gg V_{0}\, \gamma^2 N  e^{-\gamma s_c}$ one has
 \be
 A_{s} =   {  \alpha(V_{\rm up}+  V_{0})^{3}\ e^{2\sqrt{2\over 3\alpha} s_c}\over 8 \pi^{2} V_{0}^{2} }       \ .
 \ee
 and for   $ V_{\rm up}+ V_{0}   = V_{0}\, \gamma^2 N  e^{-\gamma s_c}$ one has
 \be\label{interAS}
  A_{s} =  {2(V_{\rm up}+  V_{0})   N^{2}\over 9 \alpha \pi^{2}    \   }    \ .
 \ee 
 Finally, let us calculate the tensor to scalar ratio $r$:
 \be
r =  8\left({V'\over V}\right)^{2}  = {  8V_{0}^{2} \gamma^{2}e^{ -2\gamma\,  s_N } \over  (V_{\rm up}+ V_{0})^{2} } \\ \ .
 \ee
 In the large $N$ limit one has the standard $\alpha$-attractor result
  \be
 r =    {  12\alpha  \over   N^{2}   } \ ,
 \ee
 Meanwhile for $ V_{\rm up}+ V_{0}    \gg V_{0}\, \gamma^2 N  e^{-\gamma s_c}$ the value of $r$ is  smaller, 
 \be
r =   {  8V_{0}^{2} \gamma^{2} e^{ -2\gamma\,  s_c } \over  (V_{\rm up}+ V_{0})^{2} } \ll  {  12\alpha  \over   N^{2}   }\ ,
 \ee
and for $ V_{\rm up}+ V_{0}    = V_{0}\, \gamma^2 N  e^{-\gamma s_c}$ one has
 \be\label{req}
 r =    {  3\alpha  \over   N^{2}   } \ .
 \ee
  
What is the meaning of these results?  First of all, we confirmed that in the large N limit
 \be
  N\gg {3\alpha\over 2} e^{\sqrt{2\over 3 \alpha} s_c} \left({V_{\rm up}\over V_{0}}+1\right )    \ ,
\ee 
 the predictions of $\alpha$-attractors are universal, as shown in equation \rf{pred}.  To be more precise, the amplitude of the perturbations $A_{s}$ in \rf{newampl} now depends not on $V_{0}$, but on the total height of the plateau $V_{\rm up}+  V_{0}$.
  
Meanwhile, for  smaller values of $N$  (smaller wavelengths), such that
  \be
  N\ll {3\alpha\over 2} e^{\sqrt{2\over 3 \alpha} s_c} \left({V_{\rm up}\over V_{0}}+1\right )    \ ,
\ee
which  may still exceed $N \sim 50 - 60$ for sufficiently large $V_{\rm up}$ and $\sqrt{2\over 3\alpha} s_c$, the predictions approach the flat Harrison-Zeldovich spectrum:
 \be
 A_{s} \approx   {  \alpha(V_{\rm up}+  V_{0})^{3}\ e^{2\sqrt{2\over 3\alpha} s_c}\over 8 \pi^{2} V_{0}^{2} }, \quad    
1-n_s =   \frac{4 V_{0}  e^{-\sqrt{2\over 3\alpha} s_c}}{3\alpha (  V_{\rm up} + V_{0} )   } \ll     {2\over N} ,   \quad r \approx    {  16V_{0}^{2}  e^{-2\sqrt{2\over 3\alpha} s_c} \over 3 \alpha (V_{\rm up}+ V_{0})^{2} } \ll  {  12\alpha  \over   N^{2}   }  \ .
  \ee
Note that these predictions are also universal. They do depend on constants $V_{\rm up}$,  $V_{0}$, $\alpha$ and $s_{c}$, but not on the detailed choice of the original $\alpha$-attractor potential. 

All results obtained above are formulated in terms of the field $s$ related to the field $\vp$ by the equation \rf{sf}. As we already noted, in the simplest T-model \rf{T}   one has 
$s = \vp - \sqrt{3\alpha\over 2} \ln 4  \approx \vp -1.7 \sqrt{\alpha}$.
Thus for $\alpha \lesssim 1$ one has $\vp = s +O(1)$.
  In many  cases this difference can be ignored, but if an exact relation is needed, one can always return back from $s$ to  $\vp$ in the final results using   \rf{sf}.

In particular, for the simplest hybrid inflation model \rf{hybrid}  one has
\be
 N \approx 
{3 \a (V_{\rm up}+V_{0})  \over 8 V_{0}}  \left( e^{ \sqrt{2\over 3\a}  \vp_N}
-    e^{ \sqrt{2\over 3\a}  \vp_c}\right) \ .
\ee
We have also derived this formula in Appendix A directly for the model \rf{hybrid}.

 In the limit of large $V_{\rm up}$ and/or large $\vp_{c}$ one has
 \be
 A_{s} \approx   {  \alpha(V_{\rm up}+  V_{0})^{3}\ e^{2\sqrt{2\over 3\alpha} \vp_c}\over 128 \pi^{2} V_{0}^{2} } \ , \quad    
1-n_s \approx   \frac{16 V_{0}  e^{-\sqrt{2\over 3\alpha} \vp_c}}{3\alpha (  V_{\rm up} + V_{0} )   }  \ll {2\over N}  \ ,   \quad r \approx    {  256V_{0}^{2}  e^{-2\sqrt{2\over 3\alpha}\vp_c} \over 3 \alpha (V_{\rm up}+ V_{0})^{2} } \ll  {  12\alpha  \over   N^{2}   } \ .
  \ee
  where $V_{\rm up}= {M^{4}\over 4 \lambda}$ and $V_{0} = 3 m^{2}\alpha$.

\section{Interpretation and some examples}

 Since the hybrid inflation models considered in the previous section belong to the general class of $\alpha$-attractors, some of the formal results obtained above may seem rather unexpected, especially the existence of the Harrison-Zeldovich attractor with $n_{s} = 1$. In this section we will provide a simple interpretation of our results.

The standard approach to evaluation of $n_{s}(N)$ consists of two steps. First of all, we find the point where the slow-roll approximation breaks down and inflation ends. Then we solve equations of motion to find the values of the fields driving inflation $N$ e-foldings back in the cosmological evolution, and find $n_{s}(\vp)$ at that time.

In hybrid inflation, the approach is somewhat different. We find the position of the inflaton field $\vp_{c}$ (or $s_{c}$) where the slow-roll conditions with respect to the field $\vp$ may still be satisfied, but inflation ends because of the tachyonic instability with respect to the field $\chi$. The value of the field $\vp_{c}$ depends on parameters $M$ and $g$, so by taking proper values of these parameters one can dial almost any desirable value of the field $\vp_{c}$. After that one finds $\vp_{N}$ (or, equivalently, $s_{N}$), see equation \rf{solution}.

We found that in the limit of large uplift and/or large $s_{c}$ (or $\vp_{c}$) one has $\vp_{N} \approx  \vp_{c}$ \rf{equality}. And once $\vp_{N}$ is known, one can further increase $V_{up}$ without changing $V''$. One may also exponentially decrease $V''(\vp_{N})$ by increasing $\vp_{c}$. In both cases, the slow roll parameters decrease, and $n_{s}$ asymptotically increases up to the Harrison-Zeldovich value $n_{s} = 1$.

To explain potential implications of these results, we will consider some simple numerical examples illustrating these ideas  A fully developed example of a hybrid inflation model will be considered in the next section.  

1) Let us take $\gamma = 1$, $V_{\rm up} = V_{0}$. Suppose first that we want to achieve $N=50$  e-foldings of inflation, and then trigger the waterfall transition along the lines of the hybrid inflation scenario at $s_{c}  = 1$. Then $n_{s}$ will be given by equation \rf{nsnormal}, $n_{s} = 0.96$ for $N=50$. The value of $s_{50}$ will be determined by equation \rf{solution} with $\gamma = 1$, 
\be
e^{s_{N}} =   {N\over 2}  \ .
\ee
Here we  ignored $e^{s_{0}}  \approx 2.7$ as compared to ${N\over 2} = 25$ (large $N$ approximation). This gives $s_{50} \approx \ln {25} = 3.2$.

2) Now let us change our game. Let us trigger the end of inflation not at $s_{c}  = 1$  but at $s^{*}_{c} = 3.2$. We put here a star to emphasize that this is a different regime, where inflation {\it ends} at the point $s^{*}_{c} = s_{50} \approx 3.2$. In that case (for $\gamma = 1$, $V_{\rm up} = V_{0}$) the point from which inflation goes for $N = 50$ e-foldings until it reaches $s^{*}_{c} = s_{50}= 3.2$ will be given by
\be
e^{ s^{*}_{50}} =   {50\over 2} + s^{*}_{c} = 50 \ .
\ee
Equation for $n_{s}$ for $N = 50$ will read
\be
n_{s} = 1-{1\over   N} = 0.98 \ .
\ee
 That is a significant modification of $n_{s}$ achieved by changing the point at which $N$ e-foldings of inflation end. This is achieved because if not for the waterfall, inflation from the point $ s^{*}_{N}$ would last $2N = 100$ e-foldings. We just interrupted it midway, but the calculation of $n_{s}$ for the perturbations prior to the waterfall goes the same way as if it began at the beginning of inflation of duration $N = 100$. That is why instead of $n_{s } =1-2/50$ we have $n_{s } = 1- 2/100 = 0.98$.
 
3) Let us change the game once more. Suppose that after (or during) the waterfall phase transition at $s^{*}_{c} = s_{50} \approx 3.2$ inflation does not end, but continues in the waterfall regime for  additional $\Delta N = 20$ e-foldings. This may happen, in particular,  in the models where the distance from the ridge to the minimum of the potential with respect to the field $\chi$ is greater than $M_{p}= 1$, see \cite{Garcia-Bellido:1996mdl,Clesse:2015wea} and also  a discussion in the next section near equation \rf{2inf}. Then the inflationary perturbations that we are going to see at the horizon are the ones generated in the $\alpha$-attractor regime during $N =50-\Delta N = 30$ e-foldings prior to the waterfall. This corresponds to the point from which (if not for the waterfall), the field would roll during $N = 80$ e-foldings. This yields 
\be
n_{s} =   1-{2\over 80} = 0.975 \ .
\ee
 
4) Finally, suppose that the waterfall occurs at $s_{c} = 1$. Naively, in that case one would not expect any major changes in $n_{s}$. However, this is not the case if the uplift $V_{\rm up}={M^{4}\over 4\lambda}$ is much greater than $V_{0} = 3m^{2} \alpha$. This condition  is very similar   to the standard assumption  $H^{2} = {M^{4}\over 12\lambda} \gg m^{2}$ made in the original hybrid inflation scenario \cite{Linde:1991km,Linde:1993cn}. In particular, from \rf{half} one may conclude that for $\alpha = 1$, $s_{c} = 1$,  $N = 50$  and $V_{up} \approx 11 V_{0}$ one would have 
\be
n_{s} =   1-{1\over 50} = 0.98 \ .
\ee

These examples show that a large  uplifting, or  a premature ending of the $\alpha$-attractor stage of inflation at $\gamma s_{c} \gg 1$, may lead to a significant increase of $n_{s}$ in the $\alpha$-attractor versions of the hybrid inflation models.

\section{A fully developed example}

In this section we will give a fully developed  example including all parameters of the hybrid inflation model \rf{hybrid}. In all estimates we will assume, for definiteness, that  $\alpha = 1$ (i.e. $\gamma  = \sqrt{2/3}$), the number of e-foldings is $N = 50$ and the critical value of the field is given by $s_{c} = 2$.  This corresponds to $\vp_{c} \approx  s_{c } + 1.7 = 3.7$.  In terms of the original geometric field $\phi$, the critical point is at $\phi_{c}  = 2.22$.

To evaluate the importance of the effects considered in the previous sections, we study here the intermediate regime \rf{inter}, where 
\be
n_{s} = 1-{1\over N} = 0.98 \ ,
\ee
see \rf{half}. For $\alpha = 1$   one can use \rf{req} to find 
\be
r ={3\over N^{2}}  = 0.0012 \ .
\ee
For $\alpha = 1$, $s_{c} = 2$ the condition \rf{inter} reads
\be\label{inter2}
V_{\rm up}+ V_{0}  = V_{0}\, {100\over 3}  e^{-2\sqrt{2/3}}  =  6.5 \, V_{0} \ .
\ee
Using \rf{interAS}   and Planck normalization $A_{s} = 2.1 \times 10^{{-9}}$ for  $\alpha = 1$ and $V_{0} = 3m^{2}$, we find
\be
m = 1.95 \times 10^{{-6}} \  ,
\ee
and 
\be
V_{\rm up} =   6.3  \times 10^{{-11}}\ .
\ee
Then using \rf{inter2}, we find
\be
M = 0.004\,  \lambda^{1/4} \ .
\ee
To have the critical point at $\phi_{c}  = 2.22$ one should take 
$
g = M/\phi_{c} = 0.0018  \lambda^{1/4}$.

To understand the dynamics of the waterfall instability in this model is important to compare the tachyonic mass $-M^{2}$ at $\chi = \vp = 0$ with the square of the Hubble constant at that point:
\be
H^{2}(0)={V_{\rm up}\over 3} =2.1 \times 10^{{-11}} \ .
\ee
The Hubble constant at the critical point $\phi_{c}$ is very similar. Meanwhile 
\be 
M^{2} = 1.5 \times 10^{-5} \, \sqrt \lambda \ .
\ee
Thus $M^{2} \gg H^{2}$ unless $\lambda \lesssim 10^{{-12}}$. This means that unless $\lambda$ is extremely small, the absolute value of the tachyonic mass $-M^{2}+ g^{2}\phi_{c}^{2}$ of the field $\chi$ becomes much greater than $H^{2}$  almost instantly after the inflaton field $\vp$ becomes smaller than its critical value $\vp_{c}$, and inflation ends,  just as in the original version of the hybrid inflation scenario  \cite{Linde:1991km,Linde:1993cn}.

Thus we gave here a particular example of the $\alpha$-attractor version of hybrid inflation, where $n_{s} = 1-1/N = 0.98$ instead of the standard result $n_{s} = 1-2/N = 0.96$ (for $N = 50$). This shows that by changing $V_{\rm up}$ and $\vp_{c}$ one can change $n_{s}$ anywhere in the range from $n_{s} = 1-2/N$ to $n_{s} = 1$.

This does not mean that the theory of $\alpha$-attractors is not predictive. In order to modify the standard prediction $n_{s} = 1-2/N$ we needed to consider two-field models with very special properties, such as uplifting  $V_{\rm up}$ and  a premature end of the $\alpha$-attractor stage of inflation. Nevertheless, it is important to know that such models do exist, and can be easily constructed in the familiar framework of hybrid inflation.  Other mechanisms which may lead to a premature end of inflation were reviewed 
for example in \cite{Renaux-Petel:2021yxh}.

Finally, let us try to understand what is so special about the exceptional regime $\lambda \lesssim 10^{{-12}}$. The amplitude of spontaneous symmetry breaking in the Higgs potential ${1\over 4\lambda}(M^2-\lambda\sigma^2)^2$ for $\lambda \lesssim 10^{{-12}}$ is given by
\be\label{2inf}
\sigma = M \lambda^{-1/2} \gtrsim    4 \ .
\ee
In this case, the Higgs potential ${1\over 4\lambda}(M^2-\lambda\sigma^2)^2$ becomes an inflationary potential,  because the length of the slope from $\sigma = 0$ to $\sigma = M \lambda^{-1/2}$ is super-Planckian. This length is even greater in terms of the canonically normalized field $\chi$.  It is well known that theories with super-Planckian symmetry breaking typically allow long stage of inflation, see e.g. \cite{Linde:1994hy,Vilenkin:1994pv,Linde:1994wt}.
This means that inflation may not end at the critical point, but may continue during the process of spontaneous symmetry breaking in this model.

A detailed theory of this second stage of inflation in the context of the hybrid inflation scenario is described in \cite{Garcia-Bellido:1996mdl}. The second stage of inflation may last long, or it can be short, the duration $\Delta N$ being controlled by $\lambda$. The amplitude of perturbations produced at the onset of the second stage of inflation can be very large, all the way up to $O(1)$, leading to copious formation of black holes, with masses depending exponentially on the number of e-foldings $\Delta N$ at the second stage of inflation. As proposed in  \cite{Garcia-Bellido:1996mdl,Clesse:2015wea}, primordial black holes produced in such models may be sufficiently abundant to play the role of dark matter.

The existence of the second stage of inflation means that the number of e-foldings at the $\alpha$-attractor stage is $N_{e} -\Delta N$. For example, for $N_{e} = 50$ and $\Delta N = 20$, it leaves only $N = 30$ e-foldings for $\alpha$-attractors. Then the standard  expression $n_{s} = 1-2/N$ would lead to  $n_{s} \sim 0.933$, which is ruled out by Planck2018 \cite{Planck:2018jri}.  However, in the regime  studied above one has $n_{s} = 1-1/N  \approx 0.967$, which is in a very good agreement with the Planck data.

\section{The second {\boldmath $\alpha$}-attractor regime in the same hybrid inflation model}

\begin{figure}[H]
\centering
\includegraphics[scale=0.32]{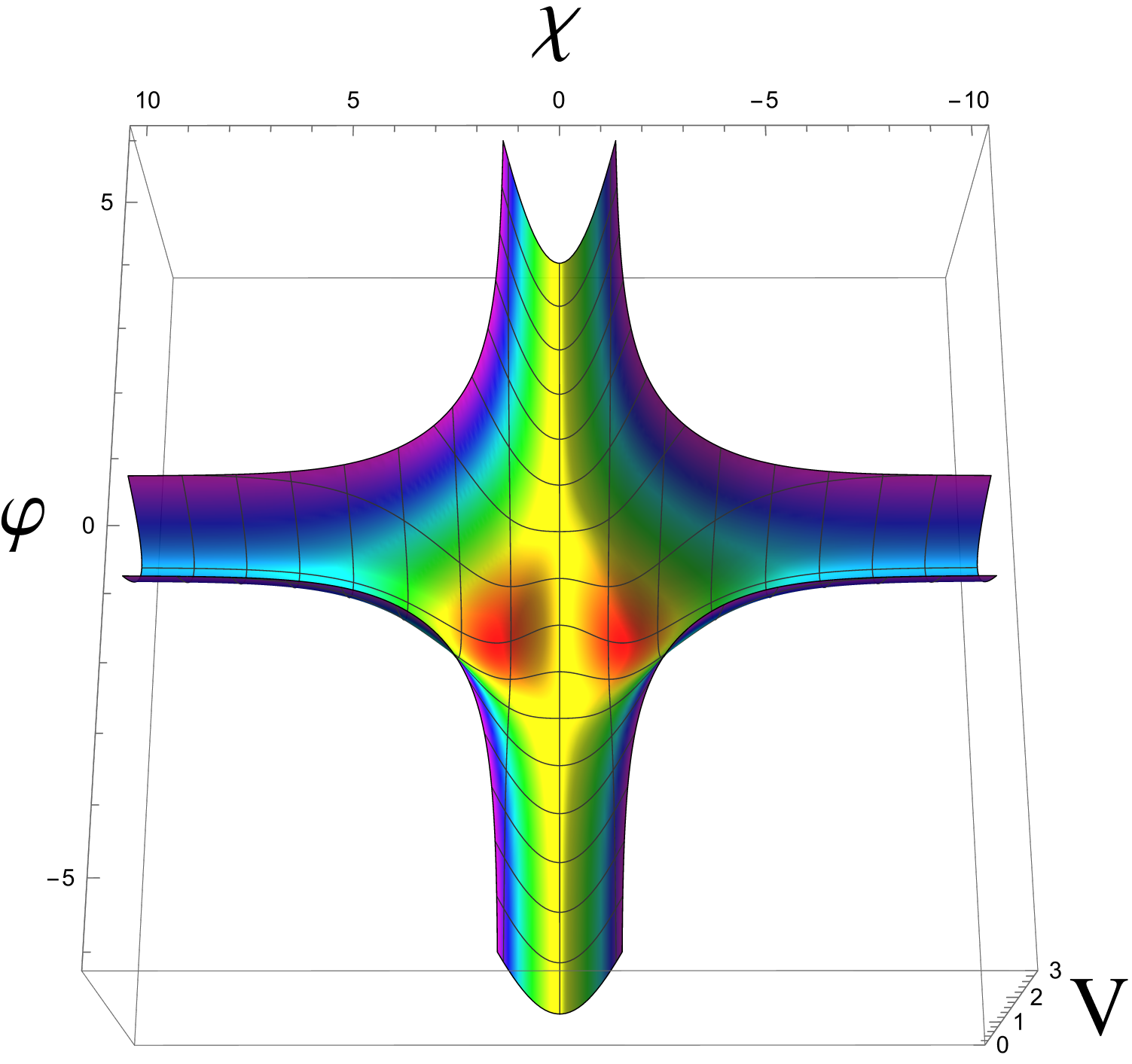}
\caption{\footnotesize  The view from the top at the hybrid inflation potential for the model \rf{hybridab} with $m = 0.2$, $M = 1$, $\lambda = 0.5$, $g = 0.8$, $\alpha = 1$, $\beta = 1$. This is the same potential as the one shown in Fig. \ref{h2}, with the same parameters,   but now we show it for a much larger range of values of $\vp$ and $\chi$.  }
\label{h22}
\end{figure}

It could seem that we already fully explored the basic hybrid inflation model  \rf{hybridab} shown in Fig. \ref{h2}. But even this simple model has some other interesting features, which are not apparent in Fig. \ref{h2}. To reveal them, we show  the potential of this model in Fig. \ref{h22}, with the same parameters as in Fig. \ref{h2}, but in a larger range of values of $\vp$ and $\chi$. 

As one can see, this potential has not one, but two flat directions, corresponding to each of the inflaton fields $\vp$ and $\chi$. Until now we studied only the scenario where the field $\vp$  rolls down along the yellow valley at $\chi = 0$, see Figs. \ref{h2} and \ref{h22}, and then the inflationary trajectory turns towards one of the two red minima of the potential at $\chi \not = 0$. All results obtained until now are describing this possibility.

The second possibility is that initially the field $\vp$ was small, whereas the field $\chi$ was large, and it was playing the role of the inflaton field, rolling down along the blue valley towards one of the two minima of its potential shown as red areas in Fig. \ref{h22}.

Fortunately, investigation of this second scenario is fairly simple. The potential of the field $\chi$ along the valley $\vp = 0$ is not uplifted by the potential of the field $\vp$, inflation ends in the standard way at the end of the slow-roll regime, so all observational consequences are described by the standard $\alpha$-attractor predictions \rf{exp}.

This means that there are two sets of cosmological predictions for the hybrid inflation model  \rf{hybridab}, depending on initial conditions for inflation. The first set corresponds to the hybrid inflation regime starting at $\chi = 0$ and large $\vp$. These predictions are described in the previous sections. The second set of predictions corresponds to the usual single-field $\alpha$-attractor regime, which begins and ends at $\vp = 0$, with the predictions given in   \rf{exp}.

\section{Hybrid polynomial attractors}\label{polS}

Similar results can be obtained for other types of plateau inflation models. Let us consider, as an example, KKLTI models with potentials approaching the plateau as inverse powers of the canonically normalized inflaton field $\vp$:
\be \label{pola}
 V \sim V_{0}(1 -{\mu^{k}\over \vp^{k}}+... ) \ ,
 \ee
where $k$ can be any (integer or not) positive constant. Such models, which were invented in the context of D-brane inflation \cite{Dvali:1998pa,Dvali:2001fw, Burgess:2001fx,Kachru:2003sx,Lorenz:2007ze,Martin:2013tda,Kallosh:2018zsi} and pole inflation scenario \cite{Galante:2014ifa, Terada:2016nqg, Karamitsos:2019vor, Kallosh:2019hzo}, were recently incorporated in the general $\alpha$-attractor framework \cite{Kallosh:2022feu}.

As before, we uplift this potential by adding to it $V_{\rm up}$, which is going to disappear after an instability at $\vp = \vp_{c}$. We will only consider here the spectral index $n_{s}$. Before the uplift, the spectral index in the large $N$ approximation is given by
\be\label{largeN}
n_{s} = 1-{2\over N} {1+k\over 2+k} \ .
\ee
After the uplift, we have
\be
n_{s} = 1-{2V_{0} k (1+k)\mu^{k} \over V_{0} k (2+k)\mu^{k}N  +(V_{\rm up } +V_{0})\, \vp_{c}^{{2+k}} } \ .
\ee
In the large $N$ limit one has the original result \rf{largeN}. In the large uplift limit (or large $\vp_{c}$ limit) one finds
\be
n_{s} = 1-{2V_{0} k (1+k)\mu^{k} \over  (V_{\rm up } +V_{0})\, \vp_{c}^{{2+k}} } \ .
\ee
In the small $k$ limit, one has the Harrison-Zeldovich result $n_{s} = 1$, 
whereas  in the intermediate case with $ (V_{\rm up } +V_{0})\, \vp_{c}^{{2+k}}  = V_{0} k (2+k)\mu^{k}N$ one has
\be\label{largeN2}
n_{s} = 1-{1\over N} {1+k\over 2+k} \ .
\ee

As in the case of exponential attractors, depending on initial conditions, there is also the standard single-field $\alpha$-attractor regime, similar to the one described in the previous section. In that case, the predictions are given by \rf{pol}.

\section{Discussion} 

In this paper we constructed $\alpha$-attractor versions of the simplest two-field hybrid inflation models. We found that the standard inflationary predictions of $\alpha$ attractors, such as $n_{s }= 1-  {2  \over  N}$, remain valid in the  limit of large number of e-foldings $N$.  However,  in some special cases the large $N$ limit is reached only beyond the horizon, for $N \gtrsim  60$, which changes predictions for the cosmological observations at $N \lesssim 60$. 

This happens because the end of inflation in the hybrid inflation scenario is not  related to breaking of the slow-roll condition for the inflaton field $\vp$, but is due to the waterfall instability with respect to the field $\chi$. Prior to the instability, which happens at $\vp < \vp_{c}$, the potential of the field $\chi$ contributes to the inflaton potential, but after the instability this contribution disappears, and inflation either ends, or continues in a very different regime. 

The critical value $\vp_{c}$  is controlled by a combination of different  parameters of the model. We studied the situations where $\vp_{c}$ belongs to the $\alpha$-attractor plateau of the potential \rf{exp} or \rf{pol}, and the universe experienced $N$ e-foldings of inflation before the field $\vp$ rolled down from $\vp_{N}$ to $\vp_{c}$. We confirmed the validity of the standard predictions of $\alpha$-attractors in the large $N$ limit. But we also found that for any particular value of $N$ there is another attractor point: In the limit of large uplift, or of large value of $\vp_{c}$, the position of the point $\vp_{N}$ moves very close to $\vp_{c}$,  all slow roll parameters become very small, and the spectral index approaches the Harrison-Zeldovich attractor point $n_{s} = 1$.

This also implies that by changing the uplifting contribution $V_{\rm up}$ of the field $\chi$, or the position of the critical point $\vp_{c}$, one can dial {\it any} desirable value of $n_{s}$ in the broad range $1-2/N \leq n_{s} \leq 1$. This does not take anything away from the universality of the standard single-field $\alpha$-attractor predictions  \rf{exp} or \rf{pol}, because this flexibility comes at a price of introducing a very specific two-field model \rf{hybrid}, \rf{cosmoAA} with  many free parameters. However, there are many situations where such flexibility can be desirable.

In this paper we only briefly outlined  some other aspects of this flexibility. In particular, now we can have a second stage of inflation during the waterfall instability without violating the observational constraints on $n_{s}$. Under some conditions (or with slight modifications of the original hybrid inflation model), this instability may  lead to production of PBHs, which may be abundant enough to play the role of dark matter \cite{Garcia-Bellido:1996mdl,Clesse:2015wea}.

 In the models with $M/g > \sqrt{6\alpha}$ the original inflationary trajectory shifts away from $\sigma = 0$, as shown in the left part of  Fig. \ref{h3}. This allows to avoid production of topological defects, while preserving most of the results obtained in this paper.
 
 Finally, there is a large spectrum of possibilities related to the potential shown in the right part of Fig. \ref{h3}.  It shows the potential for which the position of the minimum at $\sigma = M/\sqrt \lambda$ is beyond the boundary of the moduli space $\sigma = \sqrt{6\beta}$. In terms of the canonical variable $\chi$, this would mean that instead of having a minimum  at $\chi \not = 0$, we have an infinitely long plateau describing quintessence/dark energy, similar to quintessential inflation in single-field  or two-field $\a$-attractor models studied in  \cite{Dimopoulos:2017zvq,Akrami:2017cir}.  
 
 Depending on the parameters $M$ and $\lambda$, this dark energy stage may be preceded by a short waterfall stage and reheating, or a secondary inflation stage during the waterfall. For extremely small $V_{\rm up}$, one may also have a primary stage of dark energy domination during the waterfall, followed by the secondary dark energy regime during the rolling along the exponentially flat quintessential potential. Taking into account that this rolling may end up in the universe with vacuum energy that can be either positive, negative, or zero, and there can be various phase transitions along the way, modifying  density of the dark energy, we have lots of interesting possibilities to be explored.
 
We should also mention that whereas in this paper we described  hybrid inflation, some of our qualitative results may apply to other multi-field models as well, such as cascade inflation, which may occur in some string theory motivated inflationary models \cite{Kallosh:2017ced,Kallosh:2017wnt,Gunaydin:2020ric,Kallosh:2021vcf,Kallosh:2021fvz}.

\section*{Acknowledgement}
We are grateful to  Y. Yamada  for useful comments on this work and  for the collaboration on  closely related  projects  which inspired this paper.    This work is  supported by SITP and by the US National Science Foundation Grant  PHY-2014215.

\appendix

\section{Supergravity version of hybrid $\alpha$-attractors}\label{sec4}
 There are several popular versions of the hybrid inflation models in supergravity which  are known as F-term and D-term inflation \cite{Copeland:1994vg,Dvali:1994ms,Binetruy:1996xj,Halyo:1996pp}. Original versions of these models, just as the original hybrid inflation model \cite{Linde:1991km,Linde:1993cn}, required various modifications to become compatible with observations.  
 
  Cosmological $\alpha$-attractors have deep roots in supergravity describing complex fields with hyperbolic geometry  \cite{Kallosh:2013hoa,Ferrara:2013rsa,Kallosh:2013yoa,Galante:2014ifa,Kallosh:2015zsa,Kallosh:2019hzo}. In such models, kinetic terms of the scalar field   are singular at the boundary of the hyperbolic space.
 
Some of these models, the so-called E-models  \cite{Kallosh:2013yoa}, can be described  by the \K potential $K(T, \bar T) =   -3\alpha \ln (T+\bar T)$, where $T=e^{- \sqrt{2\over 3\alpha} \, \varphi (x)} +i a (x) $ is a  geometric half-plane variable. The  \K geometry $g_{T\bar T} =\partial _T \partial _{\bar T} K$ defines the relevant kinetic term ${\cal L}_{\rm kin}$ as follows:
\be
K= -3\alpha \ln (T+\bar T) \qquad \Rightarrow \qquad {\cal L}_{\rm kin}= -3\alpha {d T\ d \bar T \over (T+\bar T)^2} \ .
\ee
The kinetic term given above describes hyperbolic geometry of a half-plane $T+\bar T > 0$. The axion $a(x)$ in these models is often stabilized, and the potential depends on $t= {T+\bar T\over 2} =e^{- \sqrt{2\over 3\alpha} \varphi}$. 

The kinetic term of the scalar field $T$ is singular at the boundary $t = T+\bar T = 0$.  One may consider potentials which take the form $V= V_0 (1-t +\cdots)$ near the singularity. Then one can make a field transformation from the geometric variable $t$ to a canonically normalized field $\vp$ to reproduce  the exponential   $\a$-attractors  \rf{exp}. Potentials $V= V_0 (1-{2\over 3\alpha}  {\mu^2\over \ln^2 t}+\cdots)$ lead to polynomial $\a$-attractors  \rf{pol2}. See  \cite{Kallosh:2022feu}  for more information.
 
Similarly, one may consider the following \K\ potential of the disk variable $Z=\tanh{\vp(x)  \over \sqrt{6\alpha}} +i a (x) $:
\be
K= -3\alpha \ln (1-Z\bar Z) \qquad \Rightarrow \qquad {\cal L}_{\rm kin}= -3\alpha {d Z\ d \bar Z \over (1-Z\bar Z)^2} \ .
\ee
The kinetic term given above describes hyperbolic geometry of a Poincare disk $Z \bar Z <1$.  One may  consider  any potential $V(Z,\bar Z)$ such that the field $a$ is stabilized at $a = 0$ during inflation. If the potential is not singular  at $Z \bar Z =1$, it becomes a plateau potential $V(\tanh{\vp  \over \sqrt{6\alpha}})$ in terms of the  canonical inflaton field $\vp$ \cite{Kallosh:2013yoa}, see section \ref{sattr}. Inflationary models of such type are called  T-models  \cite{Kallosh:2013hoa}. 

 \K\ potentials mentioned above and their generalized versions often appear  in string theory related supergravity models. New powerful methods developed during the last decade allow us to construct  inflationary models in supergravity with almost any desirable potential, with any degree of supersymmetry breaking, and with any value of the cosmological constant, by using models with nilpotent fields    \cite{Ferrara:2016fwe,Kallosh:2017ced,Kallosh:2017wnt,Achucarro:2017ing,Yamada:2018nsk,Linde:2018hmx,Gunaydin:2020ric,Kallosh:2021fvz,Kallosh:2021vcf,Kallosh:2022vha}. As we will see, this includes $\alpha$-attractor models discussed in this paper.

 Here we  present two supergravity versions of the  $\alpha$-attractor generalization \rf{hybridab} of the original hybrid inflation model \rf{hybrid}. This can be done by introducing two chiral multiplets $Z_1$ and $Z_2$, both described by some hyperbolic geometries with non-canonical kinetic terms,
\be\label{ZZ}
Z_i= z_i e^{i\theta_i}
 = \tanh{\vp_i\over \sqrt{6\alpha_i}} e^{i\theta_i} \ ,
\ee
and one nilpotent multiplet $X$.

1) The first supergravity version is designed to have the angular fields $\theta_1$ and $\theta_2$ stabilized at their minimum 
$\theta_1=\theta_2=0$.
The  class of models described in  \rf{cosmoAA}, \rf{hybridab} can be presented by the following \K potential and superpotential \cite{Kallosh:2021fvz,Kallosh:2021vcf} (here we call $\vp=\vp_1, \chi=\vp_2$ and $\alpha=\alpha_1$ and $\beta=\alpha_2$). 
\be\label{KT}
K =- 3\sum_{i=1,2} \alpha_i \log  (1 - Z_i\overline{Z}_i)   +  {F_X^2 \over F_X^2 +   V_{\rm infl}(Z_i,\overline Z_i) } X\overline X,
\ee
and superpotential
\be \label{WT1}
 W=  (W_0+ F_X X) \prod_{i=1,2} (1-Z_i^{2})^{3\alpha_i/2}    \ ,
\ee
which yields
\be\label{infpot3}
V_{\rm total} (z_i) = \Lambda + V_{\rm infl}(Z_i,\overline Z_i)_{|_{Z_i=\overline Z_i = z_i}} \ ,
\ee
where $V_{\rm infl}(Z_i,\overline Z_i)$ is a  Hermitian function and $\Lambda=F_X^2 -3W_0^2$ is  the cosmological constant.  
For 
 $z_i=\tanh{\vp_i\over \sqrt{6\alpha_i}}$, $\theta_1=\theta_2=0$,  this provides a supergravity embedding of the models with a broad class of inflationary potentials of the real part of the fields $z_{i}$. In most cases,  the potentials have  stable minima at $\theta_1=\theta_2=0$, or they can be stabilized by adding some terms to the \K\ potential.

As an example, one may consider the potential 
\be \label{Zhybrid}
V_{\rm infl}(Z_i,\overline Z_i) =  3\alpha m^{2} Z_1\overline{Z _1}+{1\over 4 \lambda}(M^{2} - 6\beta \lambda Z_2\overline{Z _2})^{2} +18 g^{2 }\alpha\beta Z_1\overline{Z _1}Z_2\overline{Z _2}
\ee 
In this model the fields $\theta_i$ are stabilized, $\theta_i = 0$, and using \rf{ZZ} one can show that the potential coincides with the $\alpha$-attractor version of hybrid inflation  \rf{cosmoAA}, \rf{hybridab}. 

In this model the two inflatons are real fields. Therefore if at the end of inflation the ``Higgs'' field $\chi$ can fall to the two different minima where it has either positive or negative value, it leads to formation of domain  walls, which may lead to undesirable cosmological consequences. 

\begin{figure}[H]
\centering
\includegraphics[scale=0.45]{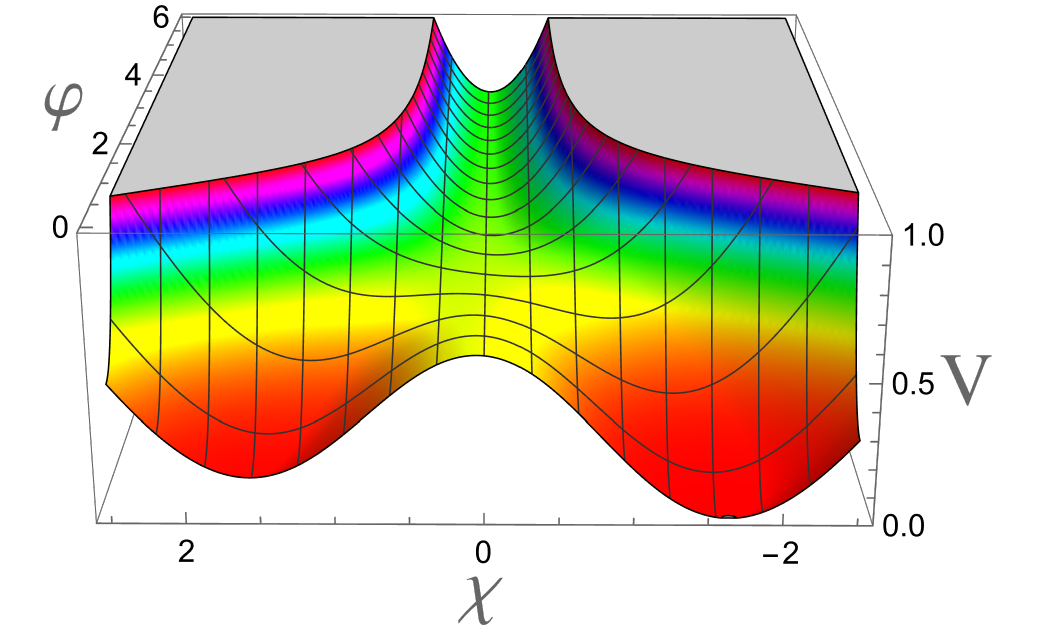}
\caption{\footnotesize  Hybrid inflation potential for the model \rf{hybridab} with $m = 0.2, M = 1, \lambda = 0.5, g = 0.8, \alpha = 1, \beta = 1$, modified by adding a small term linear in $\chi$ and by modifying $\Lambda$ to make the cosmological constant (almost exactly) vanish at the minimum. The looks very similar to the original potential shown in Fig. \ref{h2}, but inflation always ends in the minimum with $\chi < 0$. }
\label{asymm}
\end{figure}

To avoid this problem, it is sufficient to  make the potential slightly asymmetric with respect to the field $\chi$. To do it, one may add to $V_{\rm infl}(Z_i,\overline Z_i)$ a small term proportional to $  Z_2+\overline{Z _2} = 2\chi$, and also slightly modify the SUSY breaking parameter $W_0$ to achieve vanishing of the cosmological constant at the minimum of the potential. This practically does not affect the early stages of inflation,  but the term proportional to $ Z_2+\overline{Z _2} $ slightly breaks the  symmetry with respect to the change $\chi \to - \chi$, which is responsible for the formation of topological defects, see Fig. \ref{asymm}. As a result, the inflationary trajectory brings the field $\chi$ to the deeper minimum, which eliminates the domain wall problem.  

Alternatively, one may consider the version of the model in the regime shown in the left panel of Fig. \ref{h3}, where symmetry breaking occurs at the very early stages of inflation and domain walls do not form.

2) The second model of this type is a model where the complex parts of both fields are not fixed, the theory has $U(1)^{2}$ symmetry, resulting in production of cosmic strings instead of domain walls \cite{Achucarro:2017ing,Yamada:2018nsk,Linde:2018hmx,Kallosh:2022vha}. 
\bea\label{KTa}
K &=&- 3\sum_{i=1,2} \alpha_i \log  (1 - Z_i\overline{Z}_i) \nonumber  \\  &+&  {F_X^2 \, X\overline X\over \prod_{i=1,2}(1 - Z_i\overline{Z}_i)^{3 \alpha_i} \Bigl(F_X^2-3W_0^2 +   V_{\rm infl}(Z_i,\overline Z_i)\Bigr)+3W_0^2\Bigl(1-\sum_{i=1,2} \alpha_i \Bigr) } ,
\eea
and superpotential
\be \label{WT1a}
 W=   W_0+ F_X X     \ ,
\ee
For $\sum_{i=1,2} \alpha_i  < 1$ this yields 
\be\label{infpot3a}
V_{\rm total} (z_i) = \Lambda + V_{\rm infl}(Z_i,\overline Z_i)  \ ,
\ee
where $\Lambda=F_X^2 -3W_0^2$.  Importantly, this result describes the potential of the complex fields  $Z_i$, not only of their real parts as in \rf{infpot3}. This gives lots of freedom in the choice of inflationary potentials of the two fields, under the condition $\sum_{i=1,2} \alpha_i  < 1$.   

For the same choice of the hybrid inflation potential as the ones considered above in equation \rf{Zhybrid}, one reproduces the hybrid potential \rf{hybridab}, but in this context the variables $\vp$ and $\xi$  describe the absolute values of complex fields, and the potentials do not depend on the  phases $\theta_i$. For a sufficiently small amplitude of spontaneous symmetry breaking, cosmic strings produced in this scenario do not affect the amplitude of scalar perturbations. 

If one wants to avoid any topological defects, which is important if the field $\chi$ after inflation becomes large, then, just like in the previous model, one can add a small term proportional to $ Z_2+\overline{Z _2}$, or one may consider the version of the model in the regime shown in the left panel of Fig. \ref{h3}, where symmetry breaking occurs at the very early stages of inflation and cosmic strings do not form.

\section{Inflationary evolution in models  {\boldmath $V_{\rm up}+V_{0} \tanh ^2 \frac{\vp}{\sqrt{6 a}}$}}

In section \ref{predictions} we analyzed inflationary evolution in general $\alpha$-attractor models with potentials of the type
\be\label{plateau1b}
V(s) = V_{0}\large(1 -  e^{-\sqrt{2\over 3\alpha} s}+...\large) \ ,
\ee
where $s$ is given by
\be\label{sf2}
s = \vp - \sqrt{3\alpha\over 2} \ln \Bigl(2  \sqrt{6\alpha}\,{V'_{0}\over V_{0}}\Bigr)  \ ,
\ee
and  $V'_{0} = \partial_{\phi}V |_{\phi = \sqrt {6 \alpha}}$ at the boundary $\phi = \sqrt{6\alpha}$.
Here we will  do it directly in terms of the field $\vp$,  for the simplest model  
\be
V=V_{\rm up}+V_{0} \tanh ^2 \frac{\vp}{\sqrt{6 a}}  \ ,
\ee
which is a part of the hybrid inflation model \rf{hybridab}.

The number of e-foldings $N$ for  inflation beginning at the point $\vp_N$ and proceeding via slow-roll up to the point $\vp_c$ is given by
\be
 N \simeq \int ^{\vp_N}_{\vp_c}  d\vp \, {V\over V_{\vp}} \ .
\ee
Here
\be
V_{\vp}= \sqrt{\frac{2}{3\a}} V_{0} \tanh  \frac{\vp}{\sqrt{6\a}}  \text{sech}^2 \frac{\vp}{\sqrt{6\a}} \ ,
\ee
and
\be
\int   d\vp \, {V\over V_{\vp}}= {3 \a\over 4V_{0}}  \left((V_{\vp}+V_{0}) \cosh  \sqrt{{2\over 3\a} \vp} +V_{\rm up} \left(4 \log  \Bigl(\sinh  \frac{\vp}{\sqrt{6 \a}} \Bigr)-1\right)\right) \ .
\ee
Thus in the slow roll approximation 
\bea
&& N \simeq \int ^{\vp_N}_{\vp_c}  d\vp \, {V\over V_{\vp}}=   
 {3 \a\over 4V_{0}}  \left((V_{\rm up}+V_{0}) \cosh  \sqrt{{2\over 3\a} \vp_N} +4 V_{\rm up}   \log \Bigl(\sinh \frac{\vp_N}{\sqrt{6 \a}} \Bigr)\right) \cr 
- && {3 \a\over 4V_{0}}  \left((V_{\rm up}+V_{0}) \cosh  \sqrt{{2\over 3\a} \vp_c} +4V_{\rm up}   \log  \Bigl(\sinh  \frac{\vp_c}{\sqrt{6 \a}} \Bigr)\right) \ .\eea
In the $\alpha$-attractor regime with $\vp_{N} > \vp_{c}$ and ${2\over 3\a} \vp_c \gg 1$ this equation reads
\be
 N \simeq 
{3 \a (V_{\rm up}+V_{0})  \over 8 V_{0}}  \left( e^{ \sqrt{2\over 3\a}  \vp_N}
-    e^{ \sqrt{2\over 3\a}  \vp_c}\right) \ .
\label{solN}\ee
Using equation \rf{sf}, one can show that is equivalent to equation \rf{solution}, which was obtained for generic $\alpha$-attractors.

\bibliographystyle{JHEP}
\bibliography{lindekalloshrefs}
\end{document}